# Statistical Moments, Risk Weighting and Percentiles of a Loss Distribution of a Finite Homogeneous Credit Portfolio; extending the Vasicek Distribution

Marco van der Burgt [1]
*NN Group N.V.*
*P.O. Box 90504, 2509 LM The Hague, The Netherlands*
*Marco.van.der.Burgt@nn-group.com*
*November 2026*

**Abstract**
*The Vasicek distribution assumes an infinite portfolio of identical obligors and forms the basis of the Basel III/IV Risk Weighted Assets (RWA) calculation. We derive the central moments of the loss distribution for a finite number (n) of obligors. We use these moments to calculate percentiles of the fraction (x) of defaulted obligors, via the Cornish-Fisher expansion and via a weighting method. Both methods are based on scaling the standard deviation in x by a capital multiple. The weighting method assumes that the capital multiple is linear in the central moments and leads to an Extended Vasicek distribution. We apply this distribution in calculating an adjusted RWA for small portfolios. Simulations show that this adjustment is only relevant for small portfolios of less than 100 obligors and that the Extended Vasicek distribution allows analytical risk capital calculations for portfolios with different rating grades, even when the number of obligors per rating grade is small.*

[1] *The author reports no conflicts of interest. The author alone is responsible for the content and writing of the paper.*

CONTENTS

# 1 INTRODUCTION

In his seminal paper, Vasicek introduced a closed-form distribution of loan losses for credit portfolios with systemic risk expressed as a common risk factor (O. Vasicek, 1991) (O. Vasicek, 1987). This distribution, applicable for a large homogeneous credit portfolio, has been frequently applied, for example in pricing CDOs (J. Hull, 2004), importance sampling (P. Glasserman, 2003). Some authors use the Vasicek distribution as a control variate in Monte-Carlo simulations (V.Tchistiakov, 2004). The Vasicek distribution has also been extended in a multiple factor approach (J. Hull, 2004).

The Basel Committee on Banking Supervision (BCBS) also applied the Vasicek distribution in the Risk Weighted Assets (RWA) methodology. This RWA methodology first appeared in the revised Basel II framework in 2006 (BCBS, 2006). The latest final version with new requirements for the quality and quantity of regulatory capital appeared in 2011 in response to the financial crisis in 2007 – 2009 (BCBS, 2011). The Vasicek distribution is also relevant for calculating the Solvency Capital Requirement (SCR) for Type II Counterparty Default Risk in insurance (CEIOPS, 2009).

The Vasicek distribution assumes a credit portfolio of an infinite number of obligors. The credit portfolio is also assumed homogeneous, with equal weights for all exposures. Some authors extended the Vasicek model for inhomogeneous portfolios based on the Herfindahl index (M. Osadchiy, 2025). The new Basel IV proposals account for the assumptions of the Vasicek distribution by introducing a floor on the RWA. This output floor ties the output of the bank's internal risk calculation to the standardised risk approach. This prevents the bank's own internal measurement of its risk exposure from yielding less than 72.5% of the standardised approach. Basel IV is the latest step in regulatory development since the 1980s.

The aim of this paper is two-fold: first, we introduce closed-form equations for statistical moments and percentiles of the loss distribution for a portfolio with a finite number of obligors. Second, we use these results to calculate RWAs. The next section reviews the Vasicek Distribution and its properties. Section 3 introduces a theorem to calculate central moments of the loss distribution for a finite homogeneous portfolio. This theorem is applied in deriving percentiles of the loss distribution, using two methods: the Cornish-Fisher expansion and a weighting approach. Section 4 compares these methods with simulations, revealing that they result in an appropriate estimation of the 99.9% percentile. This leads to a finite portfolio adjustment of the RWA. Section 5 shows these RWA adjustments have only substantial impact when number of obligors is less than 100. The last section concludes that the current RWA with the standardized floor is sufficiently prudent, based on comparing with the finite portfolio adjustment.

## 2 REVIEW OF THE VASICEK DISTRIBUTION

Here we briefly review the Vasicek distribution (O. Vasicek, 1991). This distribution associates each obligor *i* in the portfolio with a latent variable $z_i$ that depends on a single systematic risk factor *f* and an obligor-specific factor $\varepsilon_i$ :

$$z_i = \sqrt{\rho} f + \sqrt{1-\rho}\varepsilon_i \quad f \sim \Phi[0,1] \quad \varepsilon_i \sim \Phi[0,1] \tag{1}$$

The symbol $\Phi[]$ denotes the cumulative standard normal distribution. Since $\mathbb{E}[z_i] = 0$, $\mathbb{E}[z_i^2] = 1$ and $\mathbb{E}[z_i z_j] = \rho$, Equation (1) results in a homogeneous *n x n* correlation matrix $C_n$ for a credit portfolio of *n* identical obligors:

$$C_n = \begin{bmatrix} 1 & \rho & \rho & \cdots \\ \rho & 1 & \rho & \cdots \\ \rho & \rho & 1 & \cdots \\ \vdots & \vdots & \vdots & \ddots \end{bmatrix} \tag{2}$$

When the latent variable $z_i$ is below a threshold $\tau = \Phi^{-1}[P_0]$ that is related to the unconditional Probability of Default ($P_0$), obligor *i* defaults on his debt obligations. Equation (1) shows that $z_i$ is standard normally distributed, leading to the following default probability based on condition $z_i \leq \tau$:

$$p(f) = \Phi\left[\frac{\tau - \sqrt{\rho} f}{\sqrt{1-\rho}}\right] \tag{3}$$

Then Equation (3) is applied as a conditional default probability in the cumulative binomial distribution $B_{n,p(f)}(d)$ of *d* defaults for a portfolio of *n* obligors. The portfolio is infinite and homogeneous, i.e. the exposure and default probability of every obligor is the same. The Vasicek distribution follows from integrating the binomial distribution $B_{n,p(f)}(d)$ over the systematic factor $f$ in the limit $n \to \infty$, which we call hereafter the Vasicek limit:

$$F_V(x) = \lim_{n\to\infty} P\left[\frac{k}{n} \leq x\right] = \lim_{n\to\infty} \int_0^1 B_{n,p(f)}(k)\, d\Phi(f) = \Phi\left[\frac{\sqrt{1-\rho}\,\Phi^{-1}[x] - \tau}{\sqrt{\rho}}\right] \tag{4}$$

with $x = \frac{d}{n}$ as the fraction of defaulted obligors.

We summarize here some properties of the Vasicek distribution: statistical moments, variance and percentiles, starting with the following theorem:

***Theorem 1***
*The non-central moments of the Vasicek distribution with $P_0$ as the unconditional probability of default and with correlation ρ follows from:*

$$\mathbb{E}[x^m] = \frac{1}{\sqrt{(2\pi)^m |C_m|}} \int_{-\infty}^{\tau} \cdots \int_{-\infty}^{\tau} exp\left(\frac{-\bar{u}^T C_m^{-1} \bar{u}}{2}\right) du_1 \cdots du_m = \Phi_m\left[\underbrace{\tau \dots \tau}_{m\ times}\ ; \rho\right] \quad (5)$$

*with $\tau = \Phi^{-1}[P_0]$ and matrix $C_m^{-1}$ as the inverse of the m x m homogeneous correlation matrix.*

To prove Theorem 1, we transform *x* in *u* by $F_V(x) = \Phi(u)$, giving:

$$\mathbb{E}[x^m] = \int_0^1 x^m\, dF_V(x) = \int_0^1 \left(\Phi\left[\frac{\tau - \sqrt{\rho} u}{\sqrt{1-\rho}}\right]\right)^m d\Phi(u) \quad (6)$$

Equation (29) in Appendix A shows that $\int_0^1 (\Phi[a + by])^m\, d\Phi(y) = \Phi_m\left[\underbrace{\frac{a}{\sqrt{1+b^2}} \cdots \frac{a}{\sqrt{1+b^2}}}_{m\ times}; \frac{b^2}{1+b^2}\right]$.

Substituting $a = \frac{\tau}{\sqrt{1-\rho}}$ and $b = \frac{-\sqrt{\rho}}{\sqrt{1-\rho}}$ gives Theorem 1. □

Theorem 1 gives the expectation $\mathbb{E}[x] = \Phi_1[\tau; \rho] = P_0$ for *m = 1* and the variance of the Vasicek distribution when combined with *m = 2* (C. Bluhm, 2003):

$$\sigma_{x,\infty}^2 = \mathbb{E}[x^2] - (\mathbb{E}[x])^2 = \Phi_2[\tau, \tau; \rho] - P_0^2 \quad (7)$$

with $\Phi_2[\tau, \tau; \rho] = \int_{-\infty}^{\tau} \int_{-\infty}^{\tau} \frac{1}{2\pi\sqrt{1-\rho^2}} exp\left\{-\frac{(x^2 - 2\rho xy + y^2)}{2(1-\rho^2)}\right\} dy\, dx$ as the cumulative bivariate standard normal distribution. Table 1 shows how the Vasicek moments in Theorem 1 decrease with increasing *m*.

| | *$P_0$ = 2.00%* *ρ = 0.164* | *$P_0$ = 1.00%* *ρ = 0.193* | *$P_0$ = 0.10%* *ρ = 0.234* | *$P_0$ = 0.05%* *ρ = 0.237* |
|---|---|---|---|---|
| $\Phi_1[\tau; \rho]$ | $2.00 \cdot 10^{-2}$ | $1.00 \cdot 10^{-2}$ | $1.00 \cdot 10^{-3}$ | $5.00 \cdot 10^{-4}$ |
| $\Phi_2[\tau; \rho]$ | $9.36 \cdot 10^{-4}$ | $3.26 \cdot 10^{-4}$ | $9.07 \cdot 10^{-6}$ | $2.98 \cdot 10^{-6}$ |
| $\Phi_3[\tau; \rho]$ | $7.72 \cdot 10^{-5}$ | $2.21 \cdot 10^{-5}$ | $2.78 \cdot 10^{-7}$ | $6.88 \cdot 10^{-8}$ |
| $\Phi_4[\tau; \rho]$ | $9.53 \cdot 10^{-6}$ | $2.46 \cdot 10^{-6}$ | $1.80 \cdot 10^{-8}$ | $3.60 \cdot 10^{-9}$ |
| $\Phi_5[\tau; \rho]$ | $1.58 \cdot 10^{-6}$ | $3.87 \cdot 10^{-7}$ | $1.94 \cdot 10^{-9}$ | $3.35 \cdot 10^{-10}$ |

**Table 1 Moments $\boldsymbol{\Phi_m[\tau \dots \tau; \rho]}$ of the Vasicek distribution for several values of $P_0$ and ρ.**

The percentiles at confidence level α of the Vasicek distribution follow from inverting $F_V(x_{\alpha,\infty}) = \alpha$ using Equation (4):

$$x_{\alpha,\infty} = \Phi\left[\frac{\tau + \sqrt{\rho}\Phi^{-1}[\alpha]}{\sqrt{1-\rho}}\right] \tag{8}$$

This equation is the basis of the Basel RWA calculation (BCBS, 2006). The next section introduces a modified version of Equation (8) for portfolios with a small number of obligors.

# 3 RELAXING THE VASICEK LIMIT

## 3.1 Central Moments of a loss distribution of a finite credit portfolio

Equation (7) and Equation (8) are based on approximating the cumulative binomial distribution in the Vasicek limit. Up to our knowledge, there is no closed-form equation for a finite portfolio with $n < \infty$. However, we can derive equations for the central moments of a binomial distribution with the conditional default probability as given in Equation (3). This gives an expectation $\mathbb{E}[x] = \mathbb{E}\left[\frac{d}{n}\right] = P_0$ for a portfolio of *n* obligors, of which *d* obligors default on their loans. Then the central moments are defined as $(\mu_m)_{x,n} = \mathbb{E}[(x - P_0)^m]$. By definition, the first central moment is zero, but the second central moment gives the variance in $x = \frac{d}{n}$ for a portfolio with $n < \infty$ obligors:

$$\begin{aligned} \sigma_{x,n}^2 = (\mu_2)_{x,n} &= \int_0^1 \left\{\frac{np(f) + n(n-1)p(f)^2}{n^2}\right\} d\Phi(f) - P_0^2 \\ &= \left(1 - \frac{1}{n}\right)(\Phi_2[\tau,\tau;\rho] - P_0^2) + \frac{P_0 - P_0^2}{n} \end{aligned} \tag{9}$$

in which we applied Theorem 1 in the second step. Using Equation (7), the variance $\sigma_{x,n}^2$ can be written as a $\frac{1}{n}$ - weighted average of the variance $\sigma_{x,1}^2 = P_0(1 - P_0)$ of a one-obligor portfolio and the variance $\sigma_{x,\infty}^2$ in the Vasicek limit:

$$\sigma_{x,n}^2 = \frac{\sigma_{x,1}^2}{n} + \left(1 - \frac{1}{n}\right)\sigma_{x,\infty}^2 = \frac{P_0(1 - P_0)}{n} + \left(1 - \frac{1}{n}\right)\sigma_{x,\infty}^2 \tag{10}$$

Working out central moments of order $m > 2$ shows that this $\frac{1}{n}$ - weighting also holds approximately for higher central moments. We formalize this in the following Theorem:

***Theorem 2***

*The central moments of a loss distribution with n obligors with the same unconditional default probability $P_0$ and with mutual correlation $\rho$, can be decomposed in the following:*

$$(\mu_m)_{x,n} = \frac{1}{n}(\mu_m)_{x,1} + \left(1-\frac{1}{n}\right)(\mu_m)_{x,\infty} + \frac{1}{n}\left(1-\frac{1}{n}\right)\sum_{l=1}^{m} a_{m,l}\Phi_l[\tau \ldots \tau;\rho] \qquad (11)$$

*with coefficients $a_{ml}$ summing up to zero: $\sum_{l=1}^{m} a_{m,l} = 0$ and $\Phi_l[\tau \ldots \tau;\rho]$ representing the non-central moments of the Vasicek distribution.*

We prove Theorem 2 in Appendix B. Table 2 provides the coefficients $a_{ml}$ as calculated by Equation (42) for the first central moments. The coefficients are zero for m = 2 and add up to zero for m = 3,4. The table shows that $\sum_{l=1}^{m} a_{m,l} = 0$ holds for m = 2, 3 and 4.

| | $a_{m,1}$ | $a_{m,2}$ | $a_{m,3}$ | $a_{m,4}$ | $\sum_{l=1}^{m} a_{m,l}$ |
|---|---|---|---|---|---|
| m = 2 | 0 | 0 | | | 0 |
| m = 3 | -1 | 3 | -2 | | 0 |
| m = 4 | $4P_0 - 1 - \frac{1}{n}$ | $\frac{7}{n} - 12P_0$ | $6 - \frac{12}{n} + 8P_0$ | $\frac{6}{n} - 5$ | 0 |

**Table 2 Coefficients in Equation (11) for the first three central moments m = 2, 3, 4.**

Appendix B derives the general form of the coefficients $a_{m,l}$:

$$a_{m,1} = n\sum_{k=2}^{m}\binom{m}{k}(-P_0)^{m-k}\frac{n^{-(k-2)}-1}{n-1}$$
$$a_{m,l} = n\left\{\sum_{k=l}^{m}\binom{m}{k}(-P_0)^{m-k}b_{k,l} - \binom{m}{l}(-P_0)^{m-l}\right\} \; for\; l > 1 \qquad (12)$$

in which $b_{k,l} = S_{k,l}\frac{(n-2)^{\underline{l-2}}}{n^{k-2}}$ with $S_{k,l}$ as Stirling numbers of the 2nd kind and $n^{\underline{l}}$ as the falling power $n^{\underline{l}} = n(n-1)(\ldots)(n-l+1)$ with $n^{\underline{0}} = 1$ and $n^{\underline{1}} = n$. Appendix B proves that $\sum_{l=1}^{m} a_{m,l} = 0$. Table 3 shows that coefficients $b_{kl}$ are polynomials in $\frac{1}{n}$.

| | ***$b_{k,l}$ coefficients*** |
|---|---|
| **k = 2** | $b_{2,2} = 1$ |
| **k = 3** | $b_{3,2} = \frac{3}{n},\; b_{3,3} = 1 - \frac{2}{n},$ |
| **k = 4** | $b_{4,2} = \frac{7}{n^2}, b_{4,3} = \frac{6}{n}\left(1-\frac{2}{n}\right),\; b_{4,4} = \left(1-\frac{2}{n}\right)\left(1-\frac{3}{n}\right)$ |
| **k = 5** | $b_{5,2} = \frac{15}{n^3}, b_{5,3} = \frac{25}{n^2}\left(1-\frac{2}{n}\right), b_{5,4} = \frac{10}{n}\left(1-\frac{2}{n}\right)\left(1-\frac{3}{n}\right),\; b_{5,5} = \left(1-\frac{2}{n}\right)\left(1-\frac{3}{n}\right)\left(1-\frac{4}{n}\right)$ |

**Table 3 Coefficients $b_{kl}$, introduced in Equation (32).**

## 3.2 Percentiles of a loss distribution for a finite portfolio

We may use Theorem 2 to calculate percentiles. Percentiles can be expressed in central moments like the variance, skewness and kurtosis, like the Cornish-Fisher expansion (D. Maillard, 2020). First, we introduce the *capital multiple* as a scaling factor to calculate percentiles (C. Bluhm, 2003). The capital multiple $cm_n$ expresses how much the percentile deviates from the expectation $\mathbb{E}[x] = P_0$ in terms of standard deviations:

$$x_{\alpha,n} = P_0 + cm_n \cdot \sigma_{x,n} \tag{13}$$

The standard deviation $\sigma_{x,n}$ can be calculated exactly for a portfolio of *n* obligors as the square root of the variance in Equation (10).

### 3.2.1 Capital Multiple based on Cornish-Fisher expansion

The Cornish-Fisher expansion gives the percentile $x_\alpha$ at confidence $\alpha$ (J.D. Lamb, 2016) (D. Maillard, 2020):

$$x_\alpha = \mu + \sigma\{u_\alpha + (u_\alpha^2 - 1)s + (u_\alpha^3 - 3u_\alpha)k - (2u_\alpha^3 - 5u_\alpha)s^2\} \tag{14}$$

in which $u_\alpha = \Phi^{-1}[\alpha]$, $\mu = \mathbb{E}[x]$ is the expectation value and $\sigma$ the standard deviation in *x*. Comparing (14) and (13) gives a capital multiple,

$$cm_{n,CF} = u_\alpha + (u_\alpha^2 - 1)s + (u_\alpha^3 - 3u_\alpha)k - (2u_\alpha^3 - 5u_\alpha)s^2 \tag{15}$$

Parameters *s* and *k* may be related to the skewness $S_x$ and excess kurtosis $K_x$ of *x* by $s \approx \frac{S_x}{6}$ and $k \approx \frac{K_x}{24}$ respectively. However, (J.D. Lamb, 2016) and (D. Maillard, 2020) have shown that these approximations only hold for small skewness and excess kurtosis. Generally, the skewness $S_x$ and excess kurtosis $K_x$ are also functions of parameters *s* and *k*. (J.D. Lamb, 2016) introduces a Newton-Raphson algorithm to retrieve parameters *s* and *k* from $S_x$ and $K_x$. Since both skewness and excess kurtosis are ratios of central moments, a simpler approach is to use the central moments to find *s* and *k*. Theorem 2 gives central moments $(\mu_3)_{x,n}$ and $(\mu_4)_{x,n}$ and Equation (14) gives central moments $\mu_{3,CF}(s,k)$ and $\mu_{4,CF}(s,k)$, based on Cornish-Fisher expansion:

$$\begin{aligned} \mu_{m,CF}(s,k) &= \mathbb{E}[(x-\mu)^m] \\ &= \sigma^m \mathbb{E}[(z + (z^2 - 1)s + (z^3 - 3z)k - (2z^3 - 5z)s^2)^m] \end{aligned} \tag{16}$$

with z as a standard normal distributed variable. The calculation of $\mu_{3,CF}(s,k)$ and $\mu_{4,CF}(s,k)$ using Equation (16) is cumbersome, but (J.D. Lamb, 2016) provides closed-form equations for these moments. Parameters *s* and *k* result from minimizing the following:

$$E = \left\{\mu_{3,CF}(s,k) - (\mu_3)_{x,n}\right\}^2 + \left\{\mu_{4,CF}(s,k) - (\mu_4)_{x,n}\right\}^2 , \tag{17}$$

Furthermore, Equation (14) is only valid for limited values of $s$ and $k$, since $x_\alpha$ should be a monotonously increasing function of $u_\alpha$ (D. Maillard, 2020). This requirement gives a specific domain $R$ in which parameters $s$ and $k$ are valid (J.D. Lamb, 2016):

$$R = \left\{(s,k)\middle|0 < s^2 < 3 - 2\sqrt{2}, \frac{1 + 11s^2 + \sqrt{s^4 - 6s^2 + 1}}{6} < k < \frac{1 + 11s^2 + \sqrt{s^4 - 6s^2 + 1}}{6}\right\} \tag{18}$$

The requirement $(s,k) \in R$ guarantees that $\frac{\partial x_\alpha}{\partial u_\alpha} > 0$, and therefore that $\frac{\partial cm_{n,CF}}{\partial \alpha} > 0$ by Equation (15). This imposes restrictions on applying the Cornish-Fisher approach.

### 3.2.2 Capital Multiple based on Weighting

This section provides an alternative to the Cornish-Fisher expansion to calculate percentiles $x_{\alpha,n}$ for a portfolio of n obligors. Theorem 2 shows that the central moments are approximately a $\frac{1}{n}$ - weighted average of the m[th] central moment $(\mu_m)_{x,1}$ for a one-obligor portfolio and the m[th] central moment $(\mu_m)_{x,\infty}$ in the Vasicek limit. High-order terms consist of Vasicek moments $\Phi_l[\tau \dots \tau; \rho]$, which are small for $l > 1$ as shown by Table 1, and a pre-factor $\frac{1}{n}\left(1 - \frac{1}{n}\right)$ for $m > 2$. This pre-factor is smaller than ¼ for $n \in [1, \infty)$. Assuming that the capital multiple is linear in the central moments, like Equation (14) with ignoring quadratic terms, we may approximate the capital multiple as a weighted average of a single-obligor capital multiple $cm_1$ and the capital multiple $cm_\infty$ in the Vasicek limit:

$$cm_{n,EV} \approx \frac{1}{n} cm_1 + \left(1 - \frac{1}{n}\right) cm_\infty = \frac{1}{n} cm_1 + \left(1 - \frac{1}{n}\right) \frac{x_{\alpha,\infty} - P_0}{\sigma_{x,\infty}} \tag{19}$$

with $\sigma_{x,\infty}$ and $x_{\alpha,\infty}$ as calculated from Equation (7) and (8). Equation (19) requires a capital multiple $cm_1$ for a one-obligor portfolio. The default probability density for a one-obligor portfolio is $P_0^x(1 - P_0^{1-x})$ with discrete values x = 0 or x = 1. This gives an expectation $\mathbb{E}[x] = P_0$ and a standard deviation $\sqrt{P_0(1 - P_0)}$. Since the confidence level is high, $\alpha = 99.9\%$, the obligor is certainly in default with a likelihood of 0.1%, giving $x_{\alpha,1} = 1$. As such, the capital multiple for a one-obligor portfolio equals:

$$cm_1 = \frac{x_{\alpha,1} - P_0}{\sigma_{x,1}} = \frac{1 - P_0}{\sqrt{P_0(1 - P_0)}} = \sqrt{\frac{1 - P_0}{P_0}} \tag{20}$$

Combining Equation (10), (13) and (19) gives the following equation for the percentile:

$$x_{\alpha,n} = P_0 + \left\{\frac{1}{n}\sqrt{\frac{1 - P_0}{P_0}} + \left(1 - \frac{1}{n}\right)\left(\frac{x_{\alpha,\infty} - P_0}{\sigma_{x,\infty}}\right)\right\} \cdot \sqrt{\frac{P_0(1 - P_0)}{n} + \left(1 - \frac{1}{n}\right)\sigma_{x,\infty}^2} \tag{21}$$

with $x_{\alpha,\infty} = \Phi\left[\frac{\Phi^{-1}[P_0] + \sqrt{\rho}\Phi^{-1}[\alpha]}{\sqrt{1-\rho}}\right]$ and $\sigma_{x,\infty}^2 = \Phi_2[\tau,\tau,\rho] - P_0^2$ as in Equation (7). However, $x_{\alpha,n} = 0$ when $\alpha$ is smaller than $(1 - P_0)^n$. This occurs when *n* is smaller than $\frac{ln(\alpha)}{ln(1-P_0)}$. Since $\alpha = 0.995$ in Basel III/IV, $\frac{ln(\alpha)}{ln(1-P_0)} \approx 0$ for most values of $P_0$. We will observe $x_{\alpha,n} = 0$ for small n only at low confidence levels like $\alpha = 0.9$.

We will test Equation (21) by comparing with simulations and refer to the underlying distribution $F_{EV}(x)$ with $F_{EV}(x_{\alpha,n}) = \alpha$ as the *Extended Vasicek distribution (EV)*. We verify Equation (21) by simulation in Section 4.

# 4 Testing the Vasicek Distribution and Extended Vasicek Distribution

## 4.1 Simulation set-up

We compared the Vasicek distribution and the Extended Vasicek distribution with the percentiles, based on 500,000 simulations. The objective is to see how close the percentiles of the Extended Vasicek distribution reproduce the percentiles of the simulated distribution. We performed analyses for portfolios consisting of 2 to 2000 obligors. The correlation between obligors depends from the unconditional probability of default ($P_0$), according to BCBS (BCBS, 2006) (Ong, 2005):

$$\rho = 0.12\left(\frac{1 - e^{-50 \cdot P_0}}{1 - e^{-50}}\right) + 0.24\left(1 - \frac{1 - e^{-50 \cdot P_0}}{1 - e^{-50}}\right) \tag{22}$$

Equation (22) assumes that the average correlation decreases upon increasing $P_0$, lowering the effect of the single systematic risk factor on the obligor (N. Milonas, 2024). Our test is based on $P_0$, ranging from 0.0005 to 0.02, corresponding to correlations ranging from 0.24 to 0.16 respectively. The simulations are performed in RStudio v2024.12.0, using the following steps in each simulation run:

1. A standard normal random number $z_i$ is drawn for each obligor i.

2. The random numbers $z_i$ are correlated via Cholesky decomposition of the correlation matrix as given by Equation (2).
3. An obligor defaults when $z_i \leq \Phi^{-1}[P_0]$, with $P_0$ as the unconditional probability of default of obligor i.
4. The total defaults d and fraction of defaults $x = \frac{d}{n}$ is calculated.

The important simulation outcome is the 99.9% percentile ($x_{99.9\%}^{sim}$), based on the Harrell-Davis estimator (F.E. Harrell, 1982). Furthermore, we calculated the variance, 3rd and 4th central moments of the simulations and compared with the moments as calculated by Theorem 2. We present the results as a function of the logarithmic *n*, with *n* denoting the number of obligors.

### 4.2 Comparing the Extended Vasicek distribution with simulations

First, we compare the standard deviation $\sigma_{x,sim}$ of the simulations with $\sigma_{x,\infty}$ of the Vasicek distribution as calculated by Equation (7) and with $\sigma_{x,n}$ of the finite portfolio distribution as calculated by Equation (10). Figure 1 shows that all standard deviations converge to the same value for large n, and Equation (10) gives the same values $\sigma_{x,n}$ as $\sigma_{x,sim}$ resulting from the simulations.

Theorem 2 gives theoretical values for the third central moment $(\mu_3)_{x,n}$ and fourth central moment $(\mu_4)_{x,n}$. Figure 2 and Figure 3 compare these theoretical central moments with the central moments based on the simulations, revealing that theoretical moments $(\mu_3)_{x,n}$ and $(\mu_4)_{x,n}$ agree with the corresponding simulation outcomes.

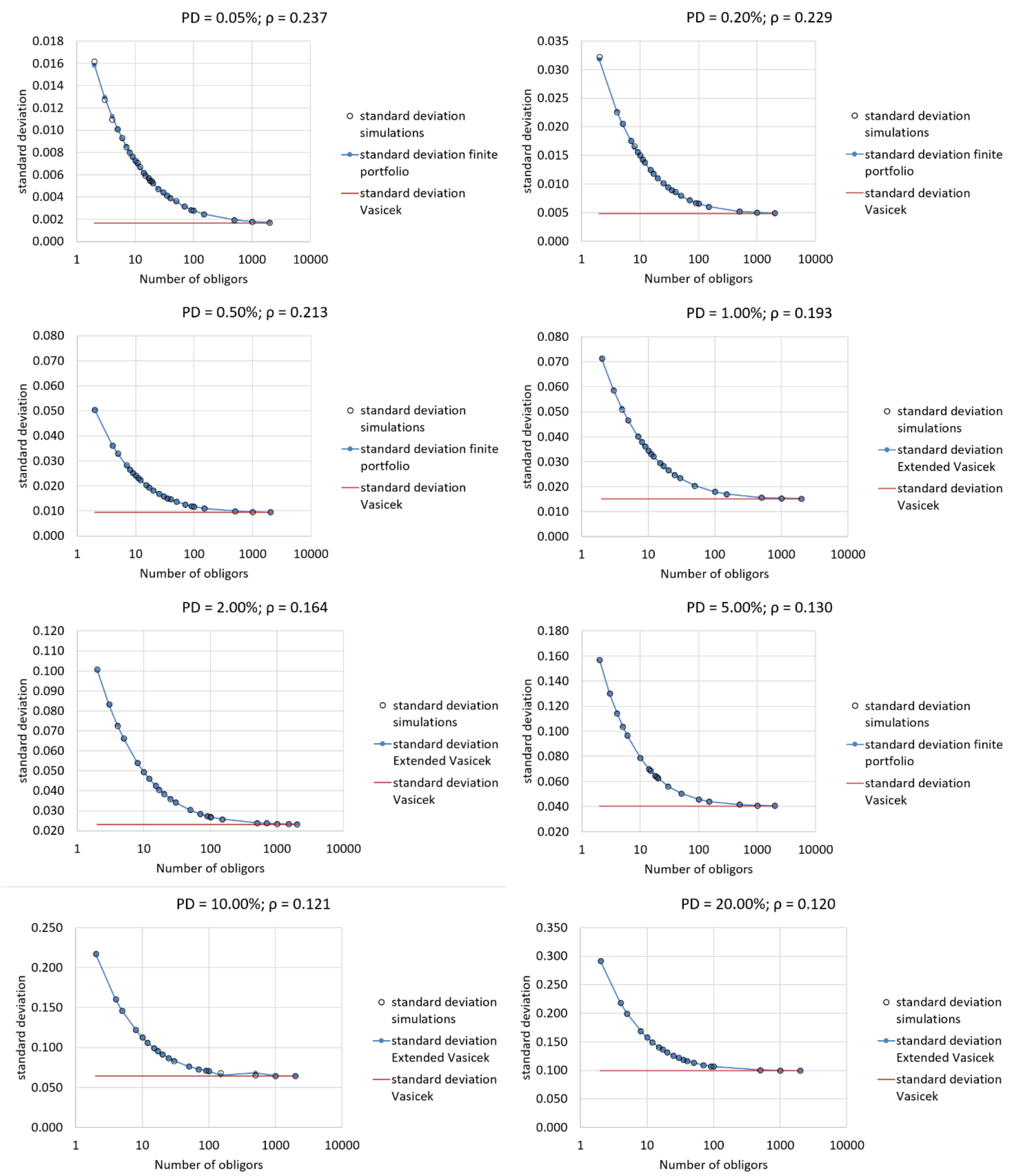

**Figure 1 The standard deviation in x ($\sigma_x$) of the Vasicek and finite portfolio distributions (n obligors), compared with $\sigma_x$ of simulations, as a function of the number of obligors in the credit portfolio.**

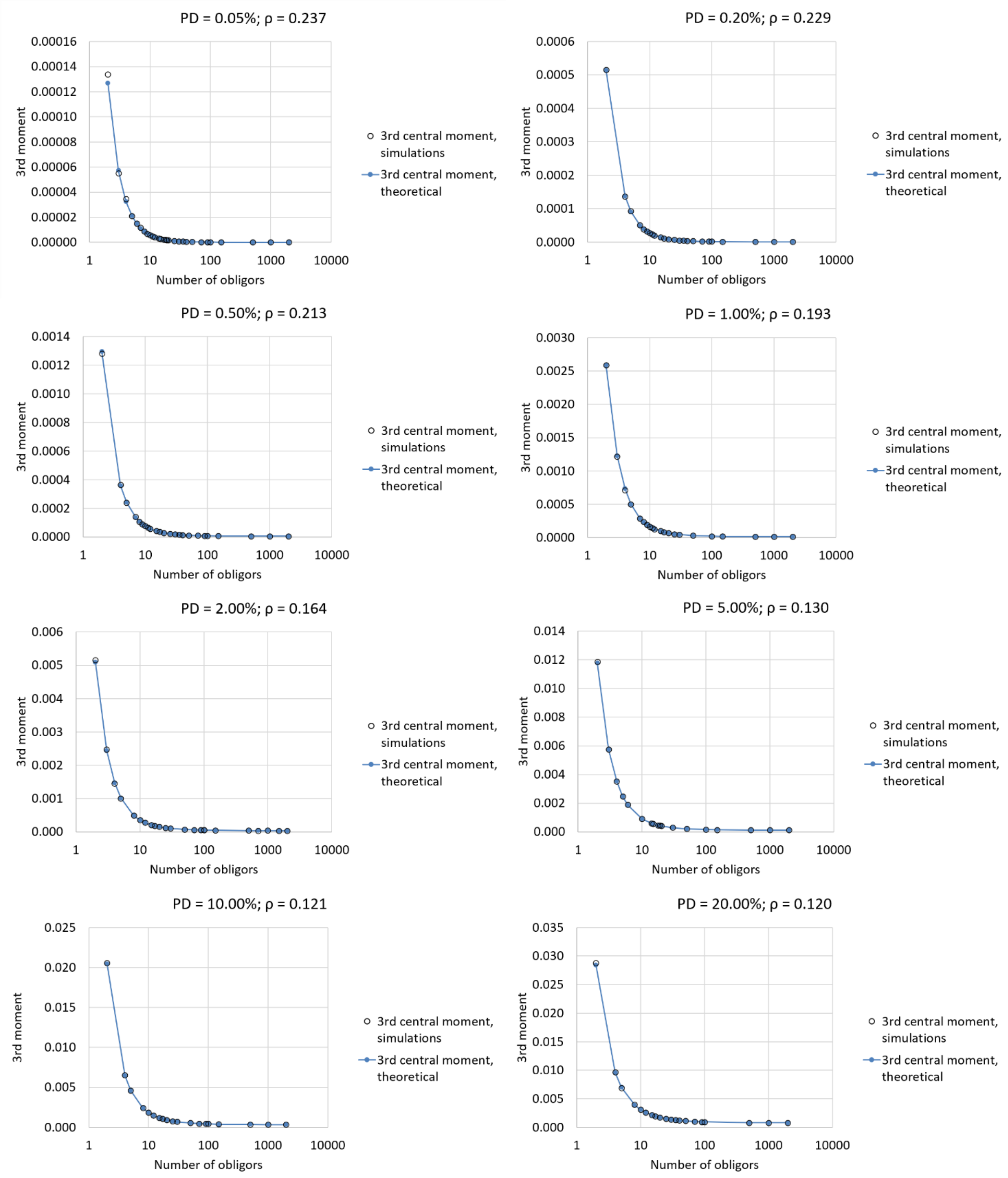

**Figure 2 The 3rd central moment in x of the finite portfolio distributions, compared with the 3rd central moment of simulations (cmoment3), as a function of the number of obligors (n) in the credit portfolio.**

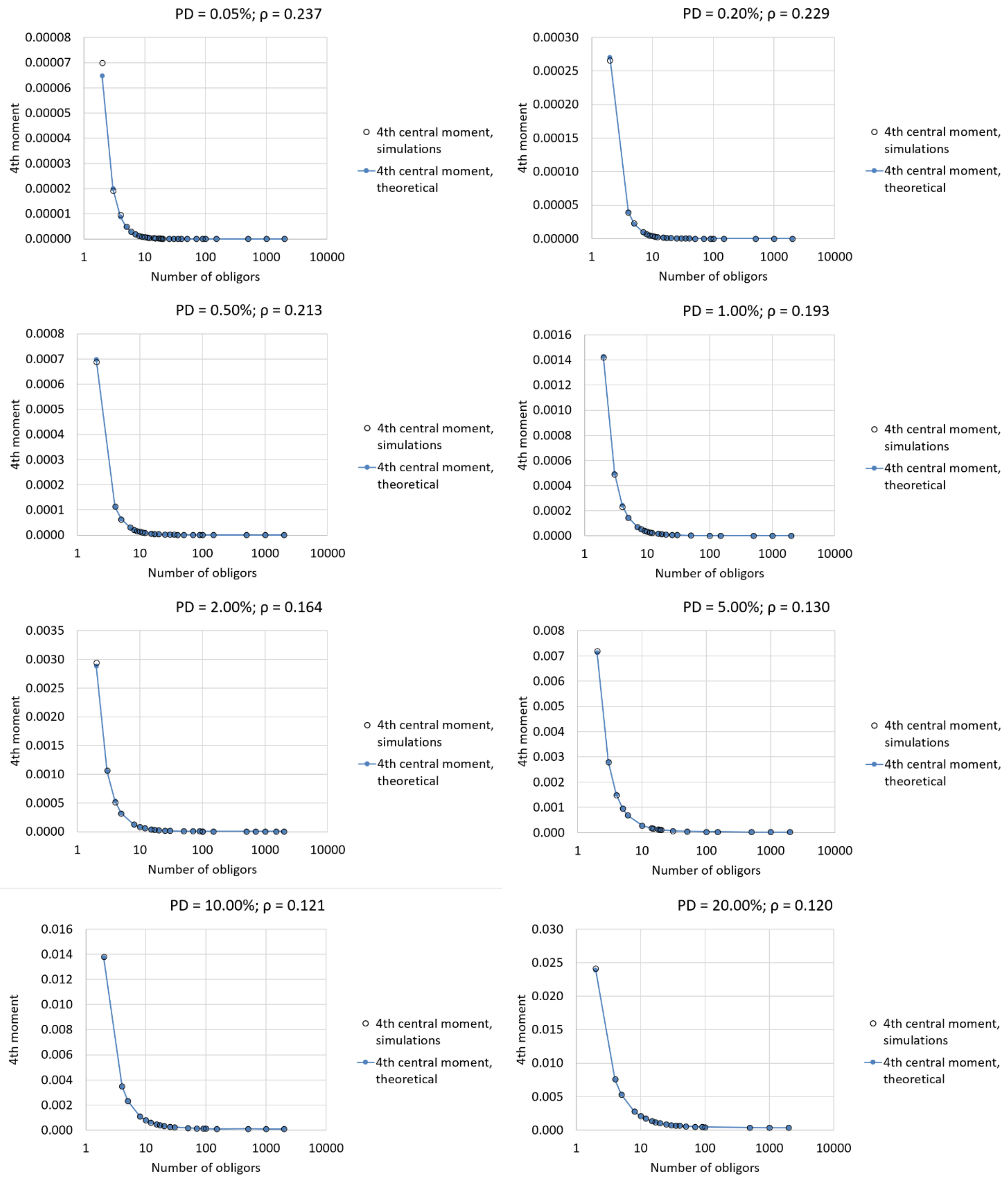

**Figure 3 The 4[rd] central moment in x of the finite portfolio distributions, compared with the 4[rd] central moment of simulations (cmoment4), as a function of the number of obligors (n) in the credit portfolio.**

*Capital Multiple based on the Cornish-Fisher expansion*

We compared the Cornish-Fisher capital multiple $cm_{n,CF}$ with the capital multiple $cm_{n,sim}$ based on simulations: $cm_{n,sim}$ follows from $\frac{x_{99.9\%}^{sim}}{\sigma_{x,n}}$ and $cm_{n,CF}$ is stepwise calculated:

1. Theorem 2 gives the central moments $(\mu_3)_{x,n}$ and $(\mu_4)_{x,n}$.
2. Minimizing *E* in Equation (17) gives *s* and *k*: a search on an $(s,k)$-grid for minimum *E* in the admissible domain *R* in Equation (18) leads to large deviations between $cm_{n,CF}$ and $cm_{n,sim}$. Minimizing *E* by the BFGS algorithm and allowing $(s,k)$ outside area *R* leads to $cm_{n,CF} \approx cm_{n,sim}$ as shown in Figure 5.
3. Use *s* and *k* in Equation (15) to calculate $cm_{n,CF}$.

Since $(s,k)$ could be outside domain *R*, Equation (15) will not always have a positive derivative, i.e. $\frac{\partial cm_{n,CF}}{\partial \alpha}$ may be negative: $cm_{n,CF}$ may not monotonously increase with increasing confidence level $\alpha$. For example, minimizing Equation (17) gives $(s = 1.42, k = 4.39) \notin R$, leading to a minimum $cm_{n,CF}$ at $\alpha = 73.8\%$. More general, Figure 4 shows $cm_{n,CF}$ for $(s,k)$ values that are determined from minimizing Equation (17). The figure demonstrates that $cm_{n,CF}$ only increases with increasing $\alpha$ for confidence levels that are larger than 90%. This works for risk capitalization at high confidence levels, but care should be taken when the Cornish-Fisher approach is applied for confidence levels lower than 90%.

Figure 5 also shows that $cm_{n,CF}$ deviates from $cm_{n,sim}$ for a small $PD \sim 10^{-4}$. Based on this observation, and that $(s,k) \notin R$ for $cm_{n,CF} \approx cm_{n,sim}$, we conclude that Cornish-Fisher expansion has limited applicability in calculating the capital multiple for RWA or risk capital calculations.

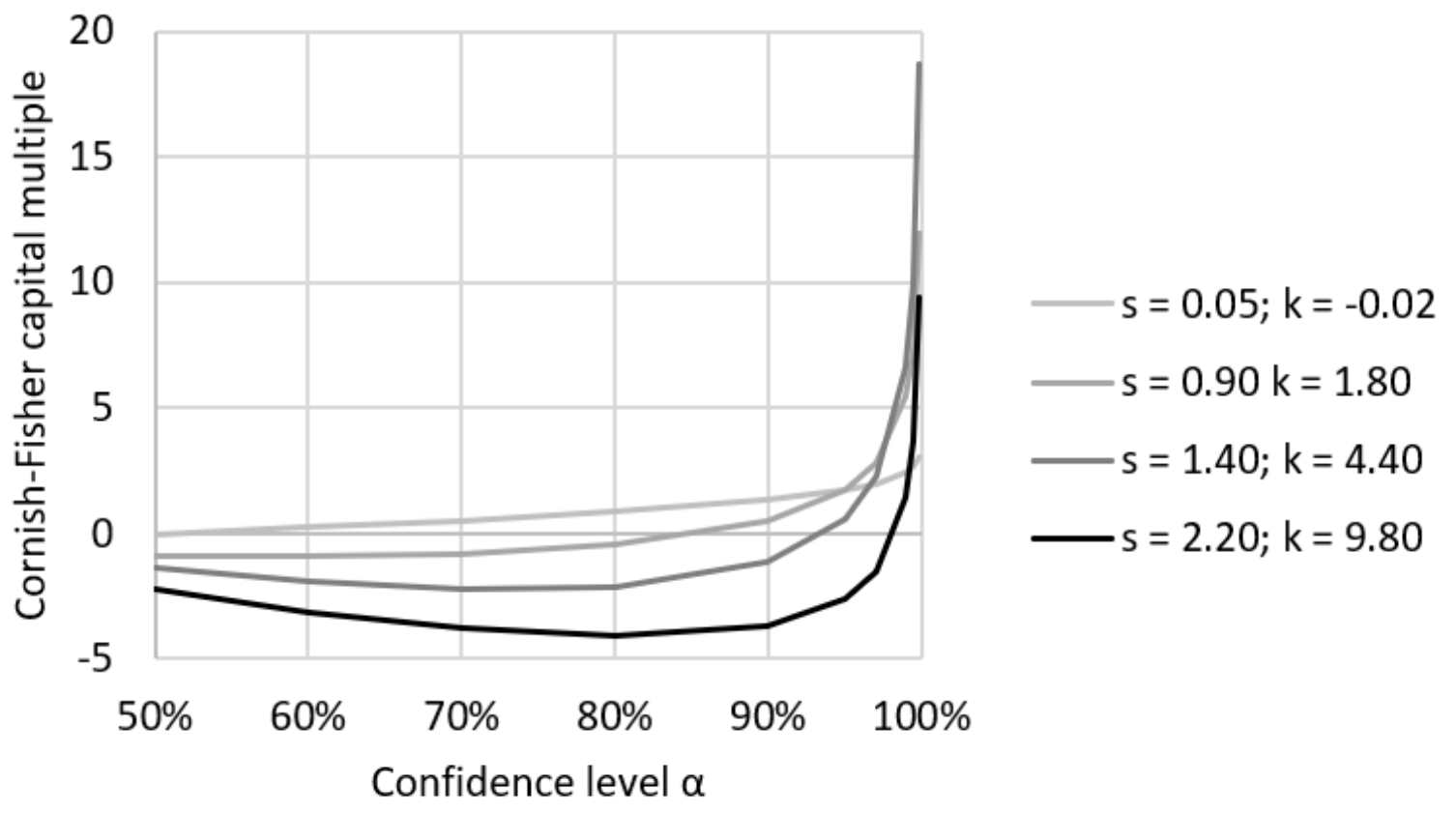

**Figure 4 The capital multiple $cm_{n,CF}$, based on Cornish-Fisher expansion, as a function of confidence level α.**

*Capital Multiple based on weighting*

The second method to calculate the capital multiple uses the approximation in Equation (19). Figure 5 also compares the capital multiple $cm_{n,sim}$ with the $cm_{n,EV}$ as calculated theoretically by Equation (19). A first observation is the choppy behaviour of $cm_{n,sim}$ for small *n*, caused by discreteness of the loss distribution as *n* becomes small. The weighting of capital multiples in Equation (19) smooths out these chops. Furthermore, both $cm_{n,sim}$ and $cm_{n,EV}$ also depend on $P_0$:

1. When $P_0 < 2\%$, both $cm_{n,sim}$ and $cm_{n,EV}$ decrease with increasing *n* and finally converge to $cm_\infty$ in Equation (19). In this region, $cm_1$ is larger than $cm_\infty = \frac{x_{\alpha,\infty} - P_0}{\sigma_{x,\infty}}$.
2. A turning point exists for $P_0 \sim 2\%$: $cm_{n,EV}$ becomes flat as a function of *n* since $cm_1 = \sqrt{(1 - P_0)/P_0}$ approaches $cm_\infty = \frac{x_{\alpha,\infty} - P_0}{\sigma_{x,\infty}}$. The difference between $cm_{n,EV}$ and $cm_{n,sim}$ relate to $\frac{1}{n}\left(1 - \frac{1}{n}\right)$ and the fourth central moment. When $cm_1 \approx cm_\infty$, high-order terms in Equation (11) become more relevant than for $cm_1 > cm_\infty$ or $cm_1 < cm_\infty$. However, this high-order effect disappears at large *n* where $cm_{n,1}$ and $cm_{n,sim}$ become equal.
3. When $P_0 > 2\%$, both $cm_{n,sim}$ and $cm_{n,EV}$ increase with increasing *n*, converging to $cm_\infty$ for high *n* values. In this region, $cm_1$ is smaller than $cm_\infty = \frac{x_{\alpha,\infty} - P_0}{\sigma_{x,\infty}}$.

To test Equation (19), we regressed $cm_{n,sim}$ against $\frac{1}{n}$ and derived $cm_1$ and $cm_\infty$ from the regression coefficients. Table 4 compares the theoretical $cm_1$ and $cm_\infty$ from Equation (19) with corresponding $cm_{1,reg}$ and $cm_{\infty,reg}$, resulting from regression. The table shows that $cm_\infty \approx cm_{\infty,reg}$, whereas $cm_1$ and $cm_{1,reg}$ show substantial differences for $P_0 < 2\%$. The slope $\left(cm_{1,reg} - cm_{\infty,reg}\right)$ of the regression is insignificant in most cases. We conclude from Figure 5 and Table 4 that $cm_1$ is hard to estimate from regression, due choppy behaviour of $cm_{n,sim}$ for small n. The $\frac{1}{n}$ - weighting smooths out this choppy behaviour in approximating the capital multiple.

| | **Theoretical, Equation (19)** | | **Regression** $cm_{n,sim} = cm_{\infty,reg} + \frac{1}{n}\left(cm_{1,reg} - cm_{\infty,reg}\right)$ | |
|---|---|---|---|---|
| **PD** | **$cm_1$** | **$cm_\infty$** | **$cm_{1,reg}$** | **$cm_{\infty,reg}$** |
| 0.05% | 44.7 | 12.1 | 36.6 | 10.7 |
| 0.20% | 22.3 | 11.1 | 17.1 | 9.6 |
| 0.50% | 14.1 | 9.9 | 10.0 | 8.8 |
| 1.00% | 9.9 | 8.7 | 5.8 | 8.2 |
| 2.00% | 7.0 | 7.4 | 3.5 | 7.0 |
| 5.00% | 4.4 | 5.8 | 6.1 | 5.6 |
| 10.00% | 3.0 | 4.9 | 3.2 | 4.8 |
| 20.00% | 2.0 | 4.0 | 2.0 | 4.0 |

**Table 4 Comparison of $cm_n$, based on Equation (19) and based on linear regression of simulated $cm_n$, versus 1/n.**

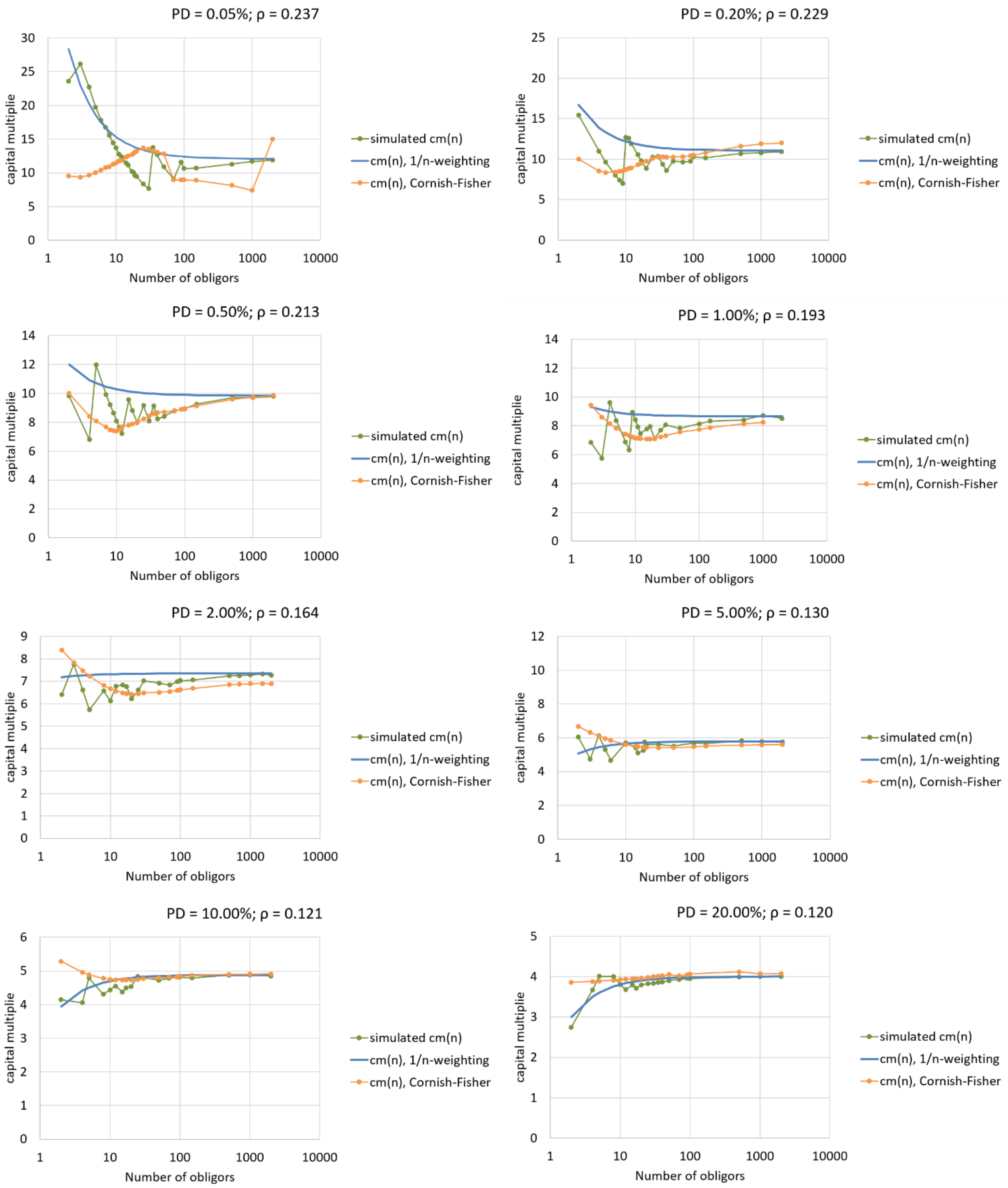

**Figure 5 The capital multiple, based on simulations (green), on the Extended Vasicek distribution for finite portfolio ($cm_{n,EV}$ , blue) and on the Cornish-Fisher expansion ($cm_{n,CF}$,orange) as a function of the number of obligors (n) in the credit portfolio.**

Figure 6 compares the percentile $x_{99.9\%}^{sim}$ of the simulations with the following:

a) The 99.9% percentile $x_{99.9\%}^{EV}$ of the Extended Vasicek distribution, based on Equation (21).

b) The 99.9% percentile $x^{V}_{99.9\%}$ of the original Vasicek distribution, based on Equation (8).

c) The 99.9% percentile $x^{CF}_{99.9\%}$, using the Cornish-Fisher expansion in Equation (15) as capital multiple.

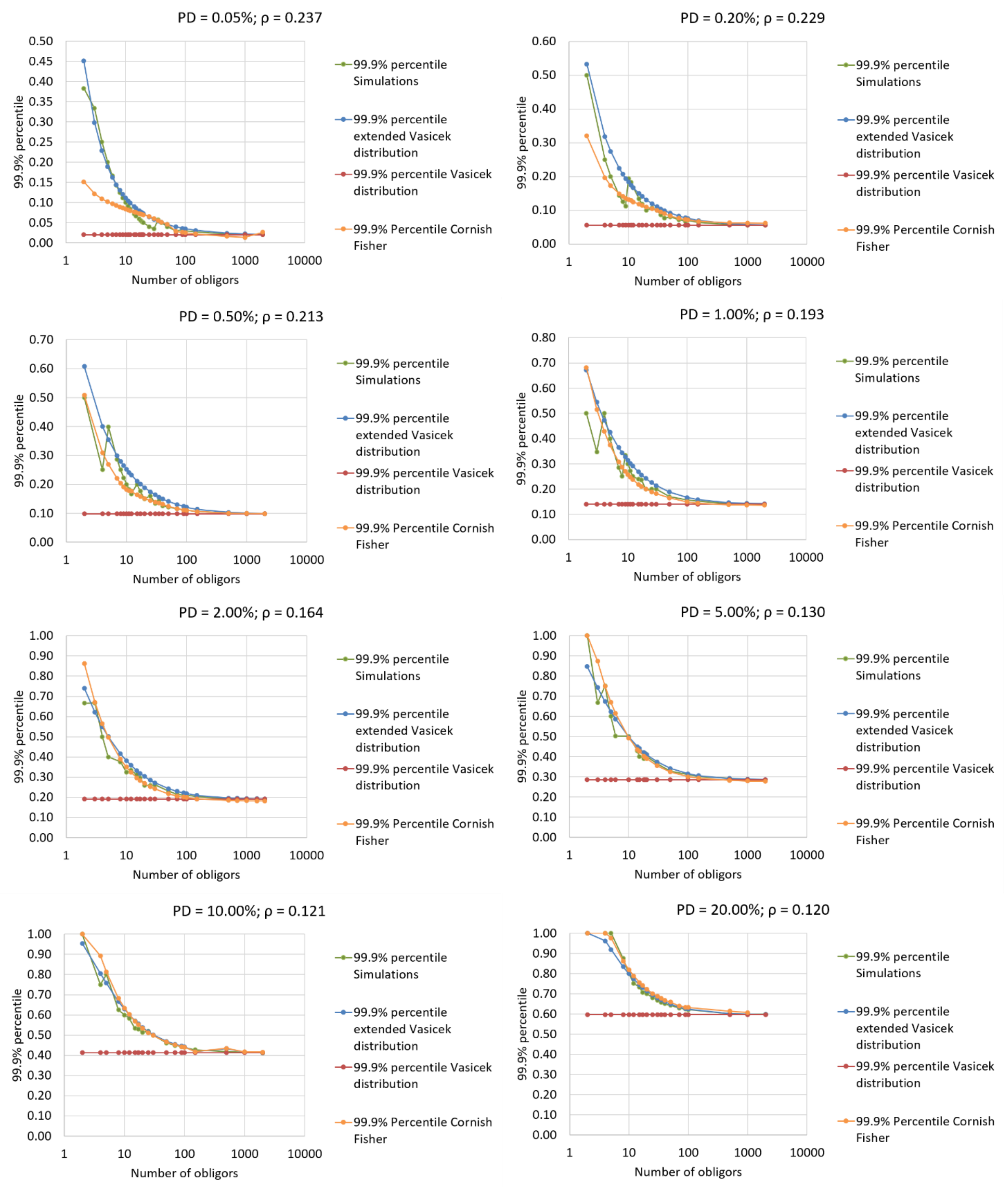

**Figure 6 The 99.9% percentiles, based on simulations (green), compared with those resulting from the Vasicek (red) distribution, the Extended Vasicek distribution (blue) and the Cornish-Fisher expansion (orange), as a function of the number of obligors (n) in the credit portfolio.**

The figure shows that the percentiles $x^{EV}_{99.9\%}$ of the Extended Vasicek distribution are considerably closer to the $x^{sim}_{99.9\%}$ than the percentiles $x^{V}_{99.9\%}$ of the original Vasicek distribution for a small number of obligors. All percentiles converge to the same limit for large n, as expected. Overall, the percentiles based on the Extended Vasicek distribution in Equation (21) agree fairly well with the simulations. We conclude that the Extended Vasicek distribution is a good approximation for credit portfolios with a finite number of obligors.

Figure 6 also compares the percentiles $x^{sim}_{99.9\%}$ with $x^{CF}_{99.9\%}$, based on the Cornish-Fisher expansion. The percentiles $x^{CF}_{99.9\%}$ align with $x^{sim}_{99.9\%}$ of the simulation, except for small $PD < 2 \cdot 10^{-3}$. Furthermore, the parameters $(s, k)$ of the Cornish-Fisher expansion are outside the admissible domain *R*.

## 5 Comparing RWAs of Vasicek and Extended Vasicek Distributions with Simulations

Equation (21) may serve as a "RWA" for a finite credit portfolio for α = 99.9%:

$$RWA_{\alpha,n} = 12.5\left\{\left[\frac{1}{n}\sqrt{\frac{1-P_0}{P_0}} + \left(1-\frac{1}{n}\right)\left(x_{\alpha,\infty} - P_0\right)\right]\sqrt{\frac{P_0(1-P_0)}{n \cdot \sigma^2_{x,\infty}} + 1 - \frac{1}{n}}\right\} \cdot LGD \cdot EAD \qquad (23)$$

with $x_{\alpha,\infty} = \Phi\left[\frac{\Phi^{-1}[P_0]+\sqrt{\rho}\Phi^{-1}[\alpha]}{\sqrt{1-\rho}}\right]$ and $\sigma^2_{x,\infty} = \Phi_2[\tau,\tau,\rho] - P_0^2$. This RWA factor converges to the current Basel RWA factor in the Vasicek limit $n \to \infty$: $RWA_{\alpha,n} = 12.5 \cdot \{x_{\alpha,\infty} - P_0\} \cdot LGD \cdot EAD$. This section compares the RWA in Equation (23) with the Basel III/IV RWA calculation.

Figure 7 shows the RWA, based on Equation (23) and the Basel III/IV RWA, floored with 72.5% of the RWA standardised approach, as a function of the number of obligors (n). We assumed an LGD of 0.4 and an EAD of 100. We consider two cases:

a) An obligor with an S&P rating A, corresponding to a PD of 0.04% and a risk weight of 50% under Basel III Standardised Approach. This gives floor of 72.5% x 50% x 100 = 36.25.
b) An obligor with an S&P rating BB, corresponding to a PD of 0.64% and a risk weight of 50% under Basel III Standardised Approach. This gives floor of 72.5% x 100% x 100 = 72.50.

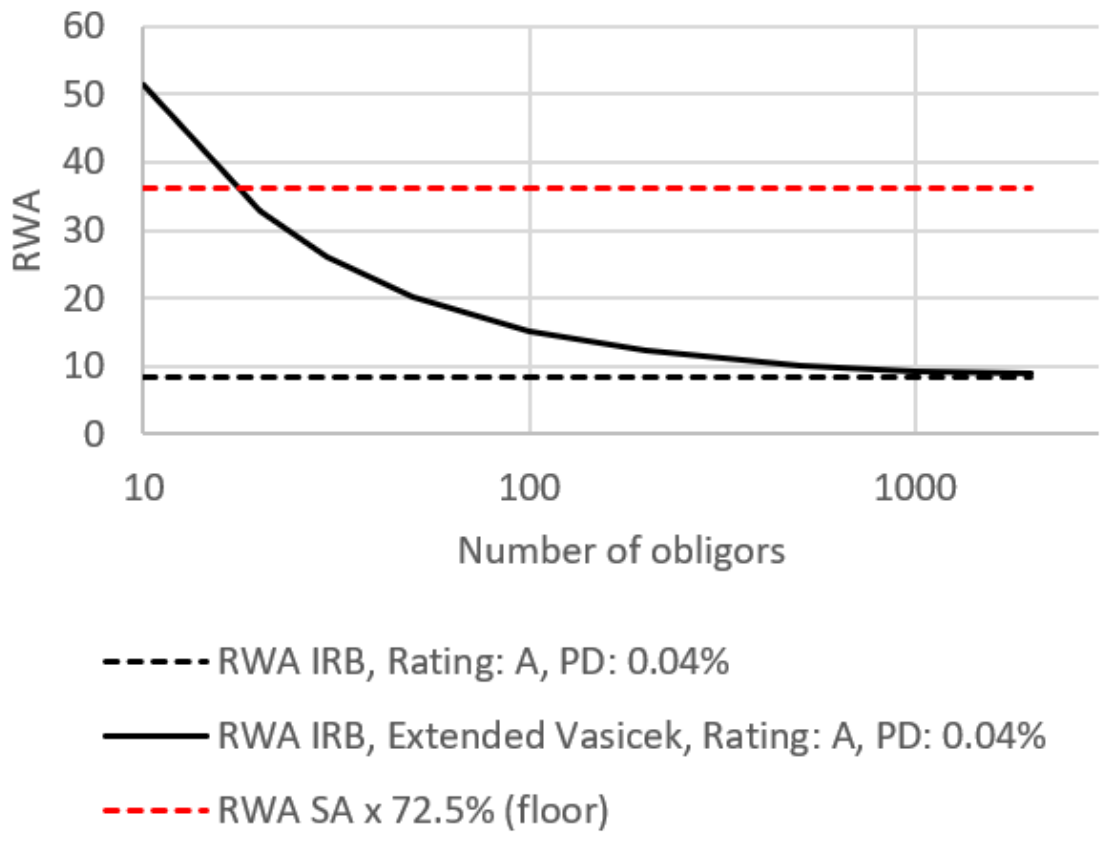

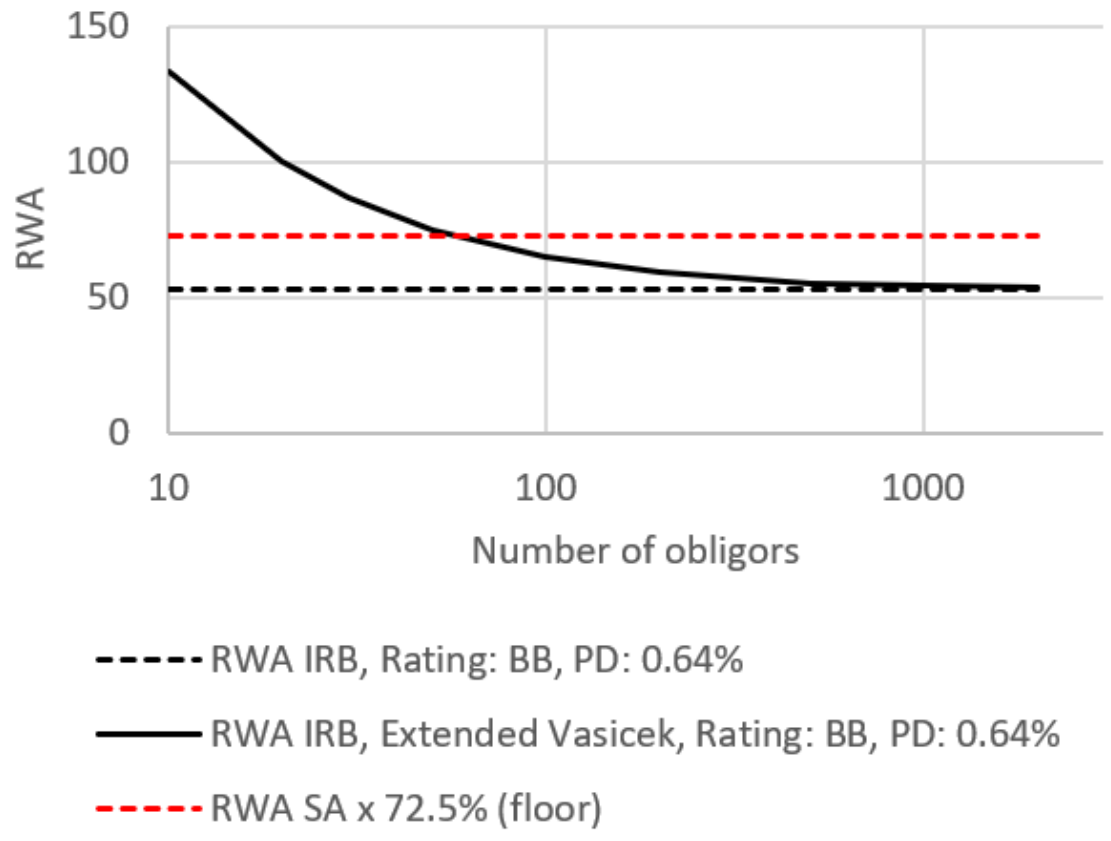

**Figure 7 Basel III/IV RWA, based on the Extended Vasicek distribution, with an LGD of 0.4, and EAD of 100 and an A and BB rated obligor.**

PDs are based on the latest observed default rates of S&P (N. Serino, 2025). The RWA based on Equation (23) converges to the Basel III/IV RWA, based on the Vasicek distribution, when *n* increases beyond 400. For small *n*, Equation (23) leads to a higher RWA than the Basel III/IV RWA, but the proposed 72.5% floor makes the Basel III/IV prudent, unless the portfolio has less than 20 obligors in case a) or less than 50 BB-rated obligors in case b).

We performed a second study for a credit portfolio, based on 4 ratings $I \in \{A, B, C, D\}$, with A representing low credit risk and D representing the highest credit risk. Obligors within the same rating category have the same PD as shown in Table 5. All obligors have the same LGD and EAD. The correlations between different rating grades result in a positive definite correlation matrix.

| Rating | PD | EAD | LGD | A | B | C | D |
|---|---|---|---|---|---|---|---|
| A | 0.001 | 100 | 0.4 | 1 | 0.25 | 0.15 | 0.10 |
| B | 0.002 | 100 | 0.4 | 0.25 | 1 | 0.17 | 0.19 |
| C | 0.015 | 100 | 0.4 | 0.15 | 0.17 | 1 | 0.22 |
| D | 0.020 | 100 | 0.4 | 0.10 | 0.19 | 0.220 | 1 |

**Table 5 Credit portfolio with 4 rating grades, their PD, EAD (in € million) and LGD and their correlations.**

We calculated RWAs for different portfolios $\Pi_1 \ldots \Pi_6$, as presented in Table 6. The first four rows present the number of obligors per rating category. Portfolio $\Pi_1$ has the same number of obligors in each grade, but Portfolio $\Pi_6$ is concentrated in rating A. The inhomogeneity is reflected by the Herfindahl–Hirschman index $HHI = \sum_{I \in \{A,B,C,D\}} f_I^2$ with $f_I$ as the fraction of obligors in rating category $I$. Row $n_{Total}$ gives the total number of obligors in the portfolio. The last three rows in Table 6 compare different RWAs for $\alpha = 99.9\%$:

- $RWA_{\alpha,BIII\backslash IV}$ is the sum of the Basel III/IV RWA per obligor. As such, this aggregation assumes perfect correlation between different rating categories. The standardized floor is not applied here.
- $RWA_{\alpha,sim}$ is calculated from 500,000 simulations as $12.5 \cdot n_{Total}\left(x_{\alpha,sim} - \langle x\rangle_{sim}\right) \cdot LGD \cdot EAD$ with $\langle x\rangle_{sim}$ as the averaged $x$ and $x_{\alpha,sim}$ as the percentile at $\alpha = 99.9\%$ resulting from the simulations.
- An analytical $RWA_{\alpha,EV}$ results from aggregating $RWA_{\alpha,n_I,I}$ in rating grade *I*, which is based on the Extended Vasicek distribution in Equation (23). Since this aggregation requires a covariance $\sigma^2_{I,J}$ between the different rating categories *I* and *J*, we use the following identity to convert the correlations $\rho_{I,J}$ in Table 5 into default correlations $\rho^{def}_{I,J}$ (C. Bluhm, 2003):

$$\sigma^2_{I,J} = \Phi_2\left[\Phi^{-1}[P_I], \Phi^{-1}\left[P_J\right]; \rho_{I,J}\right] - P_I P_J = \rho^{def}_{I,J}\sqrt{P_I(1-P_I)P_J\left(1-P_J\right)} = \rho^{def}_{I,J}\,\sigma_I\sigma_J \quad (24)$$

After multiplying each $\sigma_I$ with the capital multiple, EAD and LGD, this gives $RWA_{\alpha,EV} = n_{Total}\sqrt{\sum_{I\in\{A,B,C,D\}}\sum_{J\in\{A,B,C,D\}}\rho^{def}_{I,J} R_I R_J}$ with $R_I = f_I \cdot RWA_{\alpha,n_I,I}$.

| **Rating:** | **Π1** | **Π2** | **Π3** | **Π4** | **Π5** | **Π6** |
|---|---|---|---|---|---|---|
| A (low risk) | 10 | 6 | 5 | 100 | 200 | 260 |
| B | 10 | 6 | 5 | 8 | 10 | 10 |
| C | 10 | 21 | 5 | 8 | 8 | 26 |
| D (high risk) | 10 | 7 | 25 | 8 | 4 | 4 |
| ***$n_{Total}$*** | ***40*** | ***40*** | ***40*** | ***124*** | ***222*** | ***300*** |
| HHI | 0.25 | 0.35 | 0.44 | 0.66 | 0.82 | 0.76 |
| $RWA_{\alpha,\ BIII/IV}$ | 2052 | 2467 | 2728 | 3168 | 4541 | 6918 |
| $RWA_{\alpha,\ sim}$ | 2811 | 3266 | 3703 | 3632 | 4792 | 6638 |
| $RWA_{\alpha,\ EV}$ | 2779 | 3268 | 3649 | 3464 | 4745 | 6334 |

**Table 6 Credit portfolios ($P_i$) with 4 rating grades and different rating distributions, reflected in the HHI.**

Table 6 shows that $RWA_{\alpha,EV} \approx RWA_{\alpha,sim}$. When $n_{Total}$ is small, $RWA_{\alpha,BIII\backslash IV}$ is smaller than $RWA_{\alpha,sim}$. This aligns with expectations, as $RWA_{\alpha,BIII\backslash IV}$ is based on the Vasicek distribution, assuming large portfolios. When $n_{Total}$ increases to 200, the Vasicek distribution becomes valid and $RWA_{\alpha,BIII\backslash IV}$ approaches $RWA_{\alpha,sim}$. Portfolio $\Pi_6$ shows that $RWA_{\alpha,BIII\backslash IV}$ exceeds both $RWA_{\alpha,sim}$ and $RWA_{\alpha,EV}$, since $RWA_{\alpha,BIII\backslash IV}$ is based on adding RWAs per rating grade and does not include diversification between rating categories as reflected in Table 5.

The effect of concentration is further explored in Figure 8. The figure compares $RWA_{\alpha,sim}$ with $RWA_{\alpha,BIII\backslash IV}$ and $RWA_{\alpha,EV}$ for portfolios with different HHI and the same $n_{Total} = 40$. A high HHI indicates that obligors are concentrated in a rating category. $RWA_{\alpha,sim}$ and $RWA_{\alpha,EV}$ are similar, but $RWA_{\alpha,BIII\backslash IV}$ is lower than $RWA_{\alpha,sim}$ since $n_{Total}$ is small and the Vasicek limit does not hold. Nonetheless, all three RWAs increase with HHI, meaning that they account for concentration in a specific rating grade.

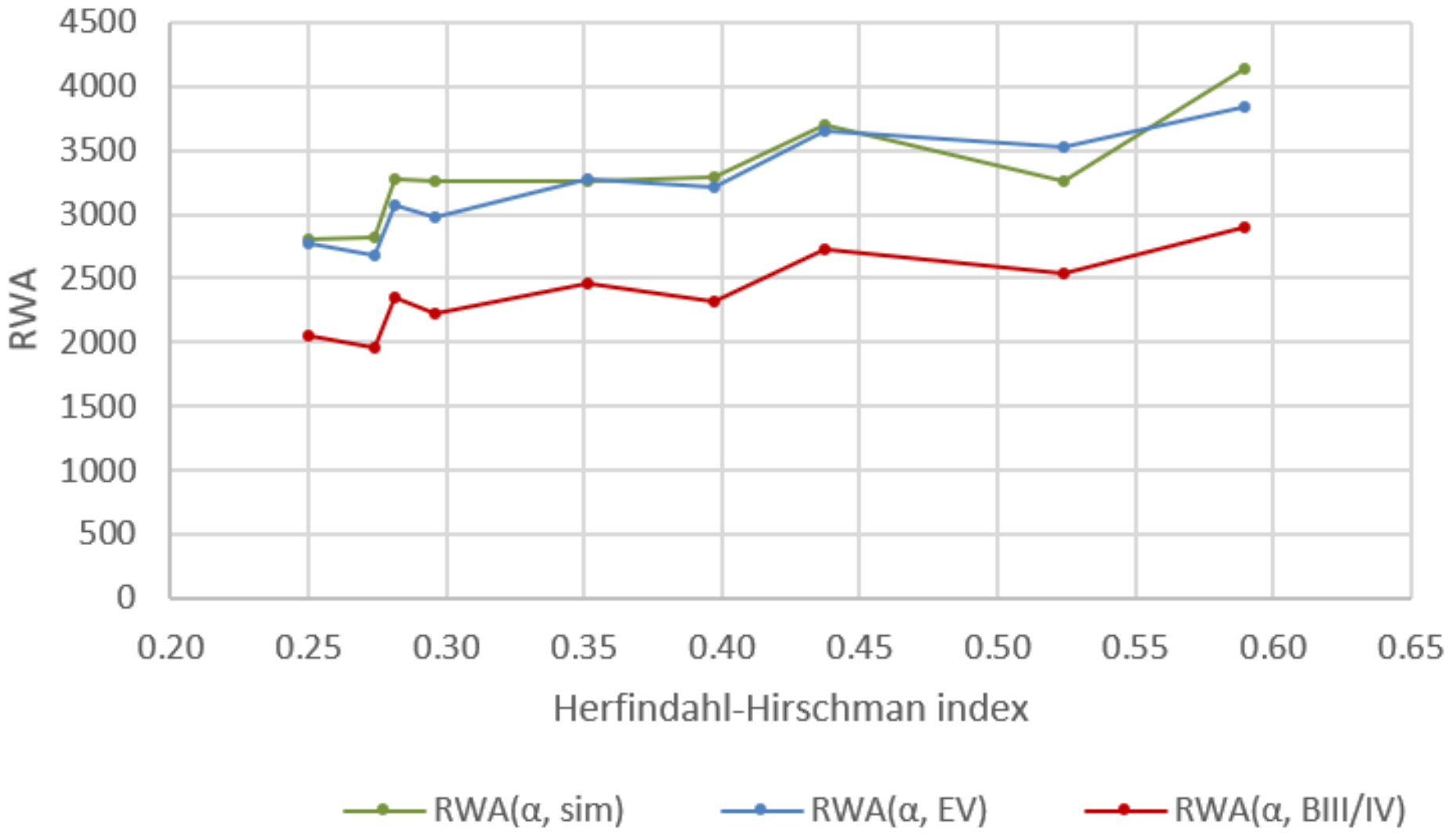

**Figure 8 RWA, resulting from simulations, Basel III/IV RWA and the RWA based on Equation (21) as a function of the HHI, assuming a portfolio of 40 obligors, distributed over four different rating grades.**

# 6 Conclusion

We introduced Theorem 2 to calculate central moments of a loss distribution of a finite credit portfolio. These central moments may be used to calculate percentiles of the loss distribution via scaling of the standard deviation with a capital multiple. We proposed two methods to achieve this: either via the Cornish-Fisher expansion or a $\frac{1}{n}$ - weighting of capital multiple and variance. We refer to the second method, giving Equation (21) for percentiles, as the Extended Vasicek distribution.

Comparison with percentiles of 500,000 simulations show that both methods work in general, but they both have drawbacks in calculating percentiles. The Cornish-Fisher method gives too low percentiles and capital multiples for small PDs. Another drawback is that the $(s, k)$-parameterization of the Cornish-Fisher expansion is outside the admissible region. As such, a capital multiple that increases monotonously with increasing confidence level is not guaranteed.

The second method gives percentiles that better align with the simulations than the Cornish-Fisher method. A drawback is the assumption that the capital multiple is linear in the central moments, so that this scaling is a $\frac{1}{n}$ – weighting of the capital multiple of a single obligor portfolio and infinite portfolio. Nonetheless, Equation (21) leads to a close approximation of the 99.9% percentiles that result from the simulations.

We deem that our analyses lead to two important conclusions. First, comparing the Basel III/IV RWA calculation with our alternative RWA, based on the Extended Vasicek distribution, shows that the Basel III/IV calculation includes enough prudency for finite portfolios, unless the portfolios are unrealistically small (n ≤ 20). When $n > 100$, the Extended Vasicek distribution approaches the original Vasicek distribution, valid for $n \to \infty$. Then the original Vasicek distribution is more convenient to apply than the Extended Vasicek distribution.

A second conclusion is that the percentiles, based on the Extended Vasicek distribution, could be used as control variate in Monte-Carlo simulations of credit portfolios. The control variate may substantially reduce variance in the Monte-Carlo calculations of percentiles (V.Tchistiakov, 2004). Simulations show that the Extended Vasicek distribution can also be applied for a portfolio with different rating grades, even when the number of obligors per rating grade is small.

## Appendix A: Calculation of integrals over the Cumulative Gaussian Distribution

This appendix evaluates the integrals over cumulative Gaussian functions. We start with the following integral:

$$\int_{0}^{1} \Phi[a+by]\,d\Phi(y) = \int_{-\infty}^{\infty}\int_{-\infty}^{a} \frac{1}{2\pi} exp\left\{-\frac{y^2}{2} - \frac{(x-by)^2}{2}\right\} dxdy$$
$$= \int_{-\infty}^{\infty}\int_{-\infty}^{a} \frac{1}{2\pi} exp\left\{-\frac{1}{2}\left[(1+b^2)\left(y-\frac{bx}{1+b^2}\right)\right]^2 - \frac{1}{2}\left(\frac{x^2}{1+b^2}\right)\right\} dxdy \quad (25)$$

in which we used completing the square in the exponential in the second step. Integration over y results in:

$$\int_{0}^{1} \Phi[a+by]\,d\Phi(y) = \int_{-\infty}^{a} \frac{exp\left\{-\frac{1}{2}\left(\frac{x^2}{1+b^2}\right)\right\}}{\sqrt{2\pi(1+b^2)}} dx = \Phi\left[\frac{a}{\sqrt{1+b^2}}\right] \quad (26)$$

The same approach may be applied in integration over Φ( ) with power m:

$$\int_0^1 (\Phi[a+by])^m \, d\Phi(y)$$
$$= \int_{-\infty}^{\infty} \underbrace{\int_{-\infty}^{a} \cdots \int_{-\infty}^{a}}_{m\ times} \frac{1}{(2\pi)^{(m+1)/2}} exp\left\{-\frac{1}{2}\left(y^2 + \sum_{i=1}^{m}(x_i - by)^2\right)\right\} \prod_{i=1}^{m} dx_1 \cdots dx_m \, dy \tag{27}$$

Completing the square in the exponential gives:

$$\int_0^1 (\Phi[a+by])^m \, d\Phi(y) =$$
$$\underbrace{\int_{-\infty}^{\infty} \int_{-\infty}^{a} \cdots \int_{-\infty}^{a}}_{m\ times} \frac{1}{(2\pi)^{(m+1)/2}} exp\left\{-\frac{1}{2}\sum_{i=1}^{m}(1+mb^2)\left(y - \frac{b\sum_{i=1}^{m} x_i}{1+mb^2}\right)^2\right. \tag{28}$$
$$\left. + \frac{1}{2}\left(\frac{(b\sum_{i=1}^{m} x_i)^2}{1+mb^2} - \sum_{i=1}^{m} x_i^2\right)\right\} dx_1 \cdots dx_m dy$$

Integration over y and collecting terms with $x_i^2$ and with $x_i x_j$ gives:

$$\int_0^1 (\Phi[a+by])^m \, d\Phi(y) =$$
$$\sqrt{\frac{(1+b^2)^m}{(2\pi)^m(1+mb^2)}} \underbrace{\int_{-\infty}^{\frac{a}{\sqrt{1+b^2}}} \cdots \int_{-\infty}^{\frac{a}{\sqrt{1+b^2}}}}_{m\ times} exp\left\{-\frac{1}{2}\left(\sum_{i=1}^{m} h_{i,i} x_i^2\right.\right. \tag{29}$$
$$\left.\left. + 2h_{i,j}\sum_{i=1}^{m}\sum_{j>i}^{m} x_i x_j\right)\right\} dx_1 \cdots dx_m$$
$$= \Phi_m\left[\frac{a}{\sqrt{1+b^2}} \cdots \frac{a}{\sqrt{1+b^2}}; \frac{b^2}{1+b^2}\right]$$

with $h_{i,i} = \frac{(1+b^2)(1+(m-1)b^2)}{1+mb^2}$ and $h_{i,j} = \frac{-b^2(1+b^2)}{1+mb^2}$. In the second step, we changed the integration interval from $(-\infty, a)$ into $\left(-\infty, \frac{a}{\sqrt{1+b^2}}\right)$.

Equation (29) results in Equation (5) when $a = \frac{\tau}{\sqrt{1-\rho}}$ and $b = \frac{\sqrt{\rho}}{\sqrt{1-\rho}}$:

- The diagonal elements and off-diagonal elements become $h_{i,i} = \frac{1+(m-2)\rho}{(1-\rho)[1+(m-1)\rho]}$ and $h_{i,j} = \frac{-\rho}{(1-\rho)[1+(m-1)\rho]}$, corresponding to the elements of the inverse correlation matrix by $C_m^{-1}$ as shown in Appendix C.
- The normalization factor becomes $\sqrt{\frac{(1+b^2)^m}{(2\pi)^m(1+mb^2)}} = \{[1+(m-1)\rho](1-\rho)^{m-1}\}^{-1/2}$, corresponding to the inverse square root of the determinant $|C_m|$ of the homogeneous correlation matrix, also shown in Appendix C.

## Appendix B: Proof of Theorem 2

To prove Theorem 2, we start with a general equation for the central moments, with $\mathbb{E}[x] = P_0$:

$$(\mu_m)_{x,n} = \mathbb{E}[(x-P_0)^m] = c_m + \sum_{k=2}^{m} c_{m,k}\mathbb{E}[x^k] \tag{30}$$

with $c_m = (-1)^m(1-m)P_0^m$ and $c_{m,k} = \binom{m}{k}(-P_0)^{m-k}$. The non-central moments $\mathbb{E}[x^k]$ in Equation (30) follow from $\mathbb{E}[d^k] = \sum_{l=0}^{k} n^{\underline{l}} S_{kl} p^l$ for a binomial distribution of *d* defaults, with the following components:

1. $S_{k,l}$ are Stirling numbers of the second kind, with $S_{k,0} = 0, S_{k,1} = S_{k,k} = 1$.
2. The falling power is defined as $n^{\underline{l}} = n(n-1)(\dots)(n-l+1)$ with $n^{\underline{0}} = 1$ and $n^{\underline{1}} = n$.

We derive central moments in terms of the fraction $x = \frac{d}{n}$ of defaulted obligors. Since the probability $p = p(f)$ is conditional on one systematic factor $f$, we use Theorem 1 for $p = p(f)$ and integrate over f:

$$\begin{aligned}\mathbb{E}[x^k] = \mathbb{E}\left[\left(\frac{d}{n}\right)^k\right] = \frac{\mathbb{E}[d^k]}{n^k} = \sum_{l=0}^{k}\frac{n^{\underline{l}}}{n^k}S_{k,l}\int_0^1 p(f)^l\, d\Phi(f) = \sum_{l=0}^{k}\frac{n^{\underline{l}}}{n^k}S_{k,l}\Phi_l[\tau\dots\tau;\rho] \\ = \frac{n}{n^k}\Phi_1[\tau;\rho] + \sum_{l=2}^{k}\frac{n^{\underline{l}}}{n^k}S_{k,l}\Phi_l[\tau\dots\tau;\rho]\end{aligned} \tag{31}$$

For a one-obligor portfolio, $n = 1$ and the non-central moments equal $\mathbb{E}[x^k] = \Phi_1[\tau;\rho] = P_0$. In the Vasicek limit, the non-central moments equal $\mathbb{E}[x^k] = \Phi_k[\tau\dots\tau;\rho]$. For the intermediate case, $1 < n < \infty$, we use $n^{\underline{l}} = n(n-1)\cdot(n-2)^{\underline{l-2}}$ and Equation (31) may be written as:

$$\mathbb{E}[x^k] = \frac{1}{n^{k-1}}\Phi_1[\tau;\rho] + \left(1 - \frac{1}{n}\right)\sum_{l=2}^{k} b_{k,l}\Phi_l[\tau\dots\tau;\rho] \quad (32)$$

with coefficients $b_{k,l} = S_{k,l}\frac{(n-2)^{\underline{l-2}}}{n^{k-2}}$. Before continuing the proof, we give some properties of coefficients and $b_{k,l}$ that will be used. Coefficients $b_{k,k}$ equal one in the Vasicek limit:

$$\lim_{n\to\infty} b_{k,k} = \lim_{n\to\infty} S_{k,k}\frac{(n-2)^{\underline{k-2}}}{n^{k-2}} = \lim_{n\to\infty}\left(1-\frac{2}{n}\right)\left(1-\frac{3}{n}\right)\dots\left(1-\frac{k-1}{n}\right) = 1 \quad k \geq 3 \quad (33)$$

and specifically, $b_{2,2} = 1$ for every n. Since $\sum_{l=0}^{k} S_{k,l}n^{\underline{l}} = n^k$ holds for Stirling numbers $S_{k,l}$, the coefficients $b_{k,l}$ add up as follows:

$$\begin{aligned}\sum_{l=2}^{k} b_{k,l} &= \sum_{l=2}^{k} S_{k,l}\frac{(n-2)^{\underline{l-2}}}{n^{k-2}} = \sum_{l=2}^{k} S_{k,l}\frac{n(n-1)\,(n-2)^{\underline{l-2}}}{n(n-1)\,n^{k-2}} = \sum_{l=2}^{k}\frac{S_{k,l}n^{\underline{l}}}{n^{k-1}(n-1)} \\ &= \frac{\sum_{l=0}^{k} S_{k,l}n^{\underline{l}} - S_{k,1}n - S_{k0}}{n^{k-1}(n-1)} = \frac{n^k - n - 0}{n^{k-1}(n-1)} = \frac{n - n^{-(k-2)}}{n-1}\end{aligned} \quad (34)$$

As shown in Table 3, the coefficients $b_{kl}$ are polynomials in $\frac{1}{n}$. We substitute Equation (32) in Equation (30):

$$(\mu_m)_{x,n} = c_m + \sum_{k=2}^{m} c_{m,k}\left\{\frac{1}{n^{k-1}}\Phi_1[\tau;\rho] + \left(1-\frac{1}{n}\right)\sum_{l=2}^{k} b_{k,l}\Phi_l[\tau\dots\tau;\rho]\right\} \quad (35)$$

Equation (35) gives the following values for an one-obligor portfolio ($n = 1$) and the Vasicek limit $n \to \infty$:

$$(\mu_m)_{x,1} = \lim_{n\to 1}(\mu_m)_{x,n} = c_m + \sum_{k=2}^{m} c_{m,k}\Phi_1[\tau;\rho] \quad (36)$$

$$(\mu_m)_{x,\infty} = \lim_{n\to\infty}(\mu_m)_{x,n} = c_m + \sum_{k=2}^{m} c_{m,k}\Phi_k[\tau\dots\tau;\rho] \quad (37)$$

in which we used Equation (33) in Equation (37). Using Equation (36) and (37) in Equation (35) gives:

$$(\mu_m)_{x,n} = \frac{1}{n}(\mu_m)_{x,1} + \left(1-\frac{1}{n}\right)(\mu_m)_{x,\infty} + R_m \quad (38)$$

with a residual term $R_m$ calculated by:

$$R_m = \frac{1}{n}\left(1-\frac{1}{n}\right)\left\{n\sum_{k=2}^{m} c_{m,k}\frac{n^{-(k-2)}-1}{n-1}\Phi_1[\tau;\rho] + n\sum_{k=2}^{m}\sum_{l=2}^{k} c_{m,k}b_{k,l}\Phi_l[\tau\ldots\tau;\rho] - n\sum_{l=2}^{m} c_{m,l}\Phi_l[\tau\ldots\tau;\rho]\right\} \quad (39)$$

Equation (39) is a linear combination of Vasicek moments $\Phi_k[\tau\ldots\tau;\rho]$ as calculated by Equation (6). Using the identity $\sum_{k=2}^{m}\sum_{l=2}^{k} x_{kl} = \sum_{l=2}^{m}\sum_{k=l}^{m} x_{k,l}$ gives:

$$R_m = \frac{1}{n}\left(1-\frac{1}{n}\right)\left\{n\sum_{k=2}^{m} c_{m,k}\frac{n^{-(k-2)}-1}{n-1}\Phi_1[\tau;\rho] + n\sum_{l=2}^{m}\sum_{k=l}^{m}\left(c_{m,k}b_{k,l}-c_{m,l}\right)\Phi_l[\tau\ldots\tau;\rho]\right\} \quad (40)$$

So we may write the residual as:

$$R_m = \frac{1}{n}\left(1-\frac{1}{n}\right)\sum_{l=1}^{m} a_{m,l}\Phi_l[\tau\ldots\tau;\rho] \quad (41)$$

with coefficients depending on $b_{kl}$ as polynomials in $\frac{1}{n}$:

$$a_{m,1} = n\sum_{k=2}^{m} c_{m,k}\frac{n^{-(k-2)}-1}{n-1}$$
$$a_{m,l} = n\left\{\sum_{k=l}^{m} c_{m,k}b_{k,l} - c_{m,l}\right\}\ for\ l > 1 \quad (42)$$

To complete the proof, we have to show that $\sum_{l=1}^{m} a_{m,l} = 0$ holds for every $m$.

$$\frac{1}{n}\sum_{l=1}^{m} a_{m,l} = \frac{1}{n}a_{m,1} + \frac{1}{n}\sum_{l=2}^{m} a_{m,l} = \sum_{k=2}^{m} c_{m,k}\frac{n^{-(k-2)}-1}{n-1} + \sum_{l=2}^{m}\sum_{k=l}^{m}\left(c_{m,k}b_{k,l}-c_{m,l}\right) \quad (43)$$

which is Equation (39) for $R_m$, divided by $\left(1-\frac{1}{n}\right)$ and with $\Phi_l[\tau;\rho]$ replaced by 1. Again, switching indices in the last term gives:

$$
\begin{aligned}
\frac{1}{n}\sum_{l=1}^{m} a_{m,l} &= \sum_{k=2}^{m} c_{m,k}\frac{n^{-(k-2)}-1}{n-1} + \sum_{k=2}^{m}\sum_{l=2}^{m} c_{m,k}b_{k,l} - \sum_{l=2}^{m} c_{m,l} \\
&= \sum_{k=2}^{m} c_{m,k}\frac{n^{-(k-2)}-1}{n-1} + \sum_{k=2}^{m}\sum_{l=2}^{m} c_{m,k}b_{k,l} - \sum_{k=2}^{m} c_{m,k} \\
&= \sum_{k=2}^{m} c_{m,k}\left(\frac{n^{-(k-2)}-1}{n-1} + \sum_{l=2}^{m} b_{k,l} - 1\right) \\
&= \sum_{k=2}^{m} c_{m,k}\left(\frac{n^{-(k-2)}-1}{n-1} + \frac{n-n^{-(k-2)}}{n-1} - 1\right) = 0
\end{aligned}
\tag{44}
$$

The last step uses Equation (34). This completes the proof of the Theorem 2. □

## APPENDIX C: INVERSE OF THE HOMOGENEOUS CORRELATION MATRIX $C_N$

The correlation matrix $C_n$ in Equation (2) has two eigenvalues:

1. The largest eigenvalue $1+(n-1)\rho$ corresponds to one eigenvector $\bar{v}_1 = [1,1,\dots,1]^T$.
2. The eigenvalue $1-\rho$ corresponds to multiple eigenvectors $\bar{v}_2 = [1,-1,0,0,0,\ \dots]^T, \bar{v}_3 = [0,1,-1,0,0,\ \dots]^T$ up to $\bar{v}_n = [\dots,0,0,0,1,-1]^T$. As such, the geometric multiplicity of this eigenvalue is n-1.

Knowing the eigenvalues and eigenvectors of $C_n$, its determinant follows from the product of the eigenvalues: $|C_n| = [1+(n-1)\rho](1-\rho)^{n-1}$ and the diagonal elements of inverse $C_n^{-1}$ follow from the minors $M_{i,i} = |C_{n-1}|$:

$$
(C_n^{-1})_{i,i} = (-1)^{i+i}\frac{M_{i,i}}{|C_n|} = \frac{|C_{n-1}|}{|C_n|} = \frac{1+(n-2)\rho}{(1-\rho)[1+(n-1)\rho]} \tag{45}
$$

Since $\bar{v}_1 = [1,1,1,\dots,1]^T$ is an eigenvector of $C_n$ with eigenvalue $1+(n-1)\rho$, it is also an eigenvector of $C_n^{-1}$ with eigenvalue $(1+(n-1)\rho)^{-1}$. Therefore, the sum of each row *i* of $C_n^{-1}$ should equal:

$$
(C_n^{-1})_{i,i} + (n-1)(C_n^{-1})_{i\neq j} = \frac{1}{1+(n-1)\rho} \tag{46}
$$

Using Equation (45) for the diagonal elements $(C_n^{-1})_{i,i}$ in Equation (46) gives:

$$
\frac{1+(n-2)\rho}{(1-\rho)[1+(n-1)\rho]} + (n-1)(C_n^{-1})_{i\neq j} = \frac{1}{1+(n-1)\rho} \tag{47}
$$

giving the off-diagonal elements $(C_n^{-1})_{i\neq j} = \frac{-\rho}{(1-\rho)[1+(n-1)\rho]}$.